\documentclass[reprint,amsmath,amssymb,aps,twocolumn]{revtex4-2}

\usepackage{dcolumn}
\usepackage{bm}
\usepackage{graphicx}
\begin{document}
\title{Fast determination of the tilt of Raman lasers using the tilt-scanned fringe for atom gravimeters}

\author{Xiaochun Duan}
\author{Wenxin Geng}
\author{Huaqing Luo}
\author{Yaoyao Xu}
 \email{xuyaoyaoleo@hust.edu.cn}
\author{Zhongkun Hu}

\affiliation{MOEKeyLaboratory of Fundamental Physical Quantities Measurement, Hubei Key Laboratory of Gravitation and Quantum Physics, PGMF, and School of Physics, Huazhong University of Science and Technology, Wuhan 430074, China} 


\affiliation{xuyaoyaoleo@hust.edu.cn}

\begin{abstract}
The sensitive axes of atom gravimeters are defined by the directions of the respective Raman lasers. Any tilt of the Raman lasers with respect to the vertical direction introduces errors in gravity measurements.
In this work, we report a fast determination of the tilt of Raman lasers, where the fringe of the atom interferometer is scanned by varying the tilt, rather than the phase, of the Raman lasers. Unlike the periodic cosine fringes typically used in atom interferometers, the fringe obtained by changing the tilt, referred to as the tilt-scanned fringe, is aperiodic and symmetric with respect to zero tilt.
The tilt-scanned fringe is highly sensitive to asymmetries caused by non-zero tilt, enabling fast and precise determination of the Raman laser tilt in atom gravimeters. 
We demonstrate that one tilt-scanned fringe, corresponding to a measurement cycle time of 13 s, can determine the tilt with a typical precision of about 30 $\mu$rad in our developed atom gravimeter. Further investigation proves that the tilt-scanned fringe approach shortens the measurement cycle time by over an order of magnitude while keeping comparable precision with conventional tilt determination techniques.
The fast tilt determination presented here is significant for the application of atom gravimeters, particularly in absolute gravity surveys.

\end{abstract}


\maketitle

\section{Introduction}
Atom interferometers have played a crucial role in high-precision absolute gravity measurements since their inception \cite{peters1999measurement}. Nowadays, atom gravimeters are recognized as a significant type of high-precision absolute gravimeters, demonstrating excellent short-term sensitivity \cite{hu2013Demonstration,Bidel2013Compactcold,Louchet-Chauvet_2011,zhang2023ultrahigh} and strong capabilities for continuous measurement with relatively high repetition rates \cite{Freier_2016,merlet2021calibration,Louchet-Chauvet_2011,Wang_2018_shiftevaluation}. 
Atom gravimeters have also shown comparable accuracies in international comparisons of absolute gravimeters alongside freely-falling corner-cube gravimeters \cite{wu2021results, Arias_2012, Francis_2015, Ullrich_2024}. Comprehensive systematic evaluations are essential for absolute gravimeters \cite{Peters_2001, Niebauer_1995}. 
For gravity surveys using atom gravimeters, such as measurements between different sites in comparative studies \cite{WU_2020} or in outdoor applications involving multiple locations \cite{zhang2021carbased}, several systematic errors must be re-evaluated on-site. 
These include errors such as effective height \cite{xu2021height}, Coriolis effect \cite{Xu_2022_evaluation, lan2012coriolis, Farah2014Effective_velocity_distribution, Wu2019gravity_surveys}, and the tilt of the sensitive axis \cite{Xu_2022_evaluation, luo2024tilt, Wu2019Orientation}. 
The rapid execution of these required on-site systematic error evaluations is critical for the practical application of atom gravimeters in gravity surveys.

Absolute gravimeters, encompassing both freely-falling corner cube types and atom interferometry types, can generally be regarded as single-axis accelerometers. The actual measured quantity of these gravimeters is the projection of the gravity acceleration vector ($\vec{g}$) along the sensitive axis. Therefore, any deviation of the sensitive axis from the vertical direction results in an error in absolute gravity measurements.
In atom gravimeters, the sensitive axis is defined by the effective wave vector of the Raman lasers. It is essential to align the Raman lasers as close as possible to the vertical direction and subsequently measure any potential tilt to evaluate the associated error \cite{peters1998}. The conventional method for determining the tilt of the Raman lasers in atom gravimeters involves intentionally altering the direction of the Raman lasers and searching for the maximum projection of $\vec{g}$ \cite{Xu_2022_evaluation,Wu2019Orientation,Menoret2018}. This maxima-search method is based on the fact that any deviation of the sensitive axis from the vertical results in a negative error. This approach can achieve a precision on the order of $\mu$rad \cite{Xu_2022_evaluation,Wu2019Orientation}. However, the method is time-consuming, as it requires scanning fringes for the Raman lasers in each configured direction to obtain the corresponding projection, and often multiple directions must be tested to locate the maximum using a parabolic fit.
In freely-falling corner cube gravimeters, the sensitive axis is also defined by the laser beam direction. Unlike atom gravimeters, the verticality of the sensitive axis in corner cube gravimeters can be conveniently tested and ensured using optical methods \cite{Niebauer_1995,Kren_2018}. Although similar optical methods have been proposed for use in atom gravimeters \cite{Senger2012A,luo2024tilt}, only a precision level of approximately 100 $\mu$rad for determining the tilt of the sensitive axis is achieved. Efforts have also been made to monitor or maintain the tilt of Raman lasers in atom gravimeters using tilt meters or other components after the tilt is determined by the maxima-search method \cite{Xie_2020,Oon2022tilt}. However, the requirement of fast and precise determination of the Raman lasers tilt isn't alleviated.

We propose an alternative method for the fast and precise determination of the tilt of Raman lasers in atom gravimeters. In this method, the direction of the Raman lasers is incrementally adjusted within one fringe, and the corresponding transition probability is recorded to form what we term a tilt-scanned fringe. In the conventional maxima-search method, the phase of the Raman lasers is scanned to create a fringe and obtain the projection of the gravity acceleration vector, $\vec{g}$, for each configured direction of the Raman lasers. In contrast, the tilt-scanned fringe method involves scanning the tilt of the Raman lasers, rather than the phase, to form the fringe.
Unlike the periodic cosine fringes produced by the maxima-search method, the tilt-scanned fringe is aperiodic, and any deviation of the Raman lasers from the vertical direction results in asymmetry in the tilt-scanned fringe. 
This property makes the tilt-scanned fringe highly sensitive to tilt, eliminating the need to obtain the $\vec{g}$ projection for tilt determination, whereas the conventional phase-scanned fringe is more sensitive to the $\vec{g}$ projection induced phase shifts. We have demonstrated this method using our developed atom gravimeter, achieving a typical precision of about 30 $\mu$rad for one tilt-scanned fringe with a measurement cycle time of 13 s. Further test of the short-term sensitivity and accuracy shows that the tilt-scanned fringe approach shortens the measurement cycle time by over an order of magnitude while keeping comparable precision with conventional tilt determination techniques.

\section{Principle}
Atom gravimeters typically use Mach-Zehnder interferometers to measure gravitational acceleration, wherein the atom wave packet is split, reflected, and recombined using Raman lasers \cite{peters1999measurement}. The coherent manipulation of the atom wave packet relies on two-photon stimulated Raman transitions \cite{Kasevich1991stimulated_transitions}, which couple the atom’s internal energy levels with its external momentum state as the wave packet evolves along two paths. The phase shift of the interferometer corresponds to the differential phase accumulated along these paths, and the resulting interferometer fringe manifests as the variation in the transition probability between two ground-state energy levels with respect to the phase shift.
The gravitational acceleration information is encoded in the fringe according to the following relation
\begin{equation}
    P = \left [1 - \cos{(\vec{k}_{\rm{eff}} \cdot \vec{g} T^2 + \Delta \varphi)}\right]/2,
\label{probability}
\end{equation}
where $P$ denotes the transition probability, $\vec{k}_{\rm{eff}}$ is the effective wave vector of the Raman lasers, $T$ is the interrogation time of the interferometer, and $\Delta \varphi$ encompasses any additional phase shifts apart from $\vec{k}_{\rm{eff}} \cdot \vec{g} T^2$. This equation indicates that the interferometer is sensitive to the projection of $\vec{g}$ along the direction of $\vec{k}_{\rm{eff}}$. Consequently, the direction of the Raman lasers defines the sensitive axis of the atom.
For accurate gravity measurements, it is crucial to align the direction of the Raman lasers as close as possible to the vertical direction. A deviation of 46 $\mu$rad in the orientation of the Raman lasers can introduce a measurement error of -1 $\mu$Gal.

\begin{figure}
    \centering
    \includegraphics[width=0.75\linewidth]{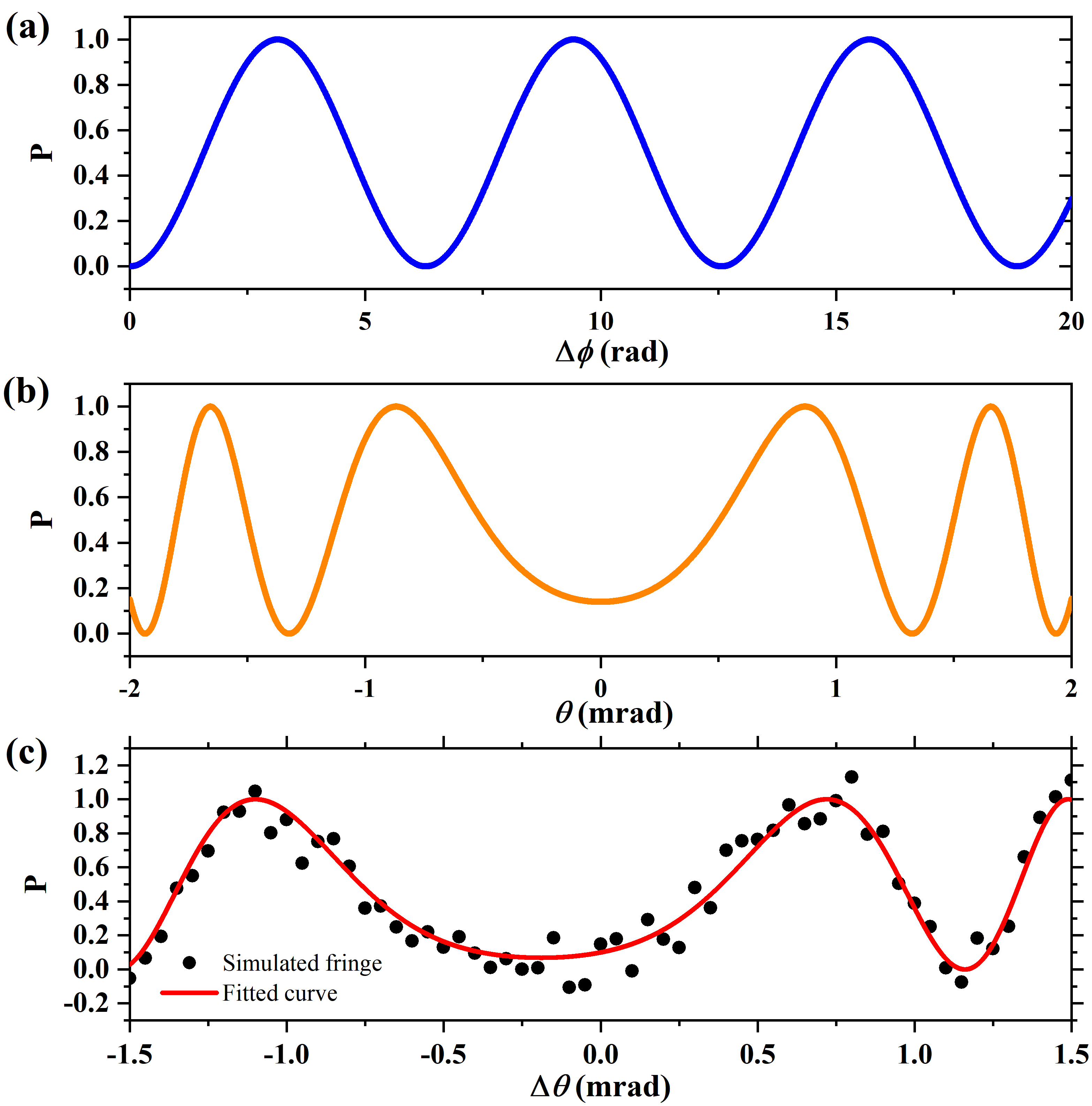}
    \caption{Different fringes manifested as the variation of the transition probability $P$ in atom interferometers. (a) The variation of $P$ is induced by the phase change of the Raman lasers. This phase-scanned fringe is a periodic cosine fringe, typically used for gravity measurements by atom interferometers. (b) The variation of $P$ is induced by the tilt change of the Raman lasers. This tilt-scanned fringe is aperiodic and symmetric with respect to zero tilt. (c) A numerical simulation of the tilt-scanned fringe with $\theta_0=$200 $\mu$rad and a noise component (with a root-mean-square value of 0.1) added to $P$. The red line represents the fit line, yielding $\theta_0^{\rm{fit}}=$190(6) $\mu$rad.}
    \label{fringe}
\end{figure}

For the precise determination of the tilt of Raman lasers, the usual approach involves changing the direction of the effective wave vector, $\vec{k}_{\rm{eff}}$, and measuring the corresponding projection of the gravitational acceleration vector, $\vec{g}$. Once the maximum projection of $\vec{g}$ is identified, the reference for the vertical direction is established, allowing the initial tilt of the Raman lasers to be determined. In this so-called maxima-search method, for each direction of $\vec{k}_{\rm{eff}}$, the phase of the Raman lasers is varied to induce changes in the transition probability, forming a cosine fringe, as illustrated in Fig. \ref{fringe} (a). In practice, the frequency chirp rate of the Raman lasers is typically adjusted to scan the fringe. This adjustment is considered as an equivalent phase change of the Raman lasers, as the fringe for gravity measurements is conventionally plotted as the variation of transition probability against the phase rather than the chirp rate. The $\vec{g}$ projection is then determined by performing a cosine fit to this measured phase-scanned fringe (generally requiring tens of shots 
measurements within one fringe). 
Once projections in various directions of the Raman lasers are obtained, a parabolic fit is applied to the projection versus direction data to find the maximum projection and establish the corresponding vertical reference. Although the maxima-search method can achieve a precision on the order of $\mu$rad, it is notably time-consuming.

Alternatively, we propose scanning the fringe by varying the tilt. The expression for the fringe in Eq. (\ref{probability}) can be modified to explicitly reflect the dependence on the tilt of the Raman lasers
\begin{equation}
    P=\left \{1-\cos{\left [k_{\rm{eff}}g T^2 \cos{(\theta_0+\Delta \theta)}+\Delta \varphi\right ]}\right \}/2,
\label{Probability_theta}
\end{equation}
where $\theta_0$ represents the initial tilt of the Raman lasers, and $\Delta \theta$ denotes the intentional variation of the tilt. The variation in transition probability as a function of $\Delta \theta$ forms the so-called tilt-scanned fringe, as illustrated in Fig. \ref{fringe} (b) (the horizontal is represented by $\theta=\theta_0+\Delta\theta$). 
Unlike the phase-scanned fringe, the tilt-scanned fringe is aperiodic and centered at $\theta_0 = 0$. Any non-zero initial tilt ($\theta_0 \neq 0$) results in a shift of the tilt-scanned fringe and introduces asymmetry with respect to $\Delta \theta$. Therefore, the tilt-scanned fringe is expected to be highly sensitive to the initial tilt $\theta_0$.

To assess the sensitivity of the tilt-scanned fringe, a simulated fringe with an initial tilt of $\theta_0 = 200$ $\mu$rad is shown in Fig. \ref{fringe} (c). This simulation is based on Eq. (\ref{Probability_theta}), where $\Delta \theta$ varies from -1.5 mrad to 1.5 mrad in steps of 0.05 mrad, and a noise component with a root-mean-square value of 0.1 is added to the calculated transition probability. The simulated fringe is fitted using the function $P = A + B \cos{[k_{\rm{eff}} g T^2 \cos{(\theta_0 + \Delta \theta)} + \Delta \varphi]}$, where the parameters to be determined are the offset $A$, amplitude $B$, phase shift $\Delta \varphi$, and tilt $\theta_0$. The fitting process yielded $\theta_0 = 190(6)$ $\mu$rad, demonstrating the sensitivity of the tilt-scanned fringe to the initial tilt.
Notably, high precision prior knowledge of the gravitational acceleration $g$ is not required for the tilt determination here. A similar fitting procedure with $g$ as an additional free parameter resulted in $\theta_0 = 202(6)$ $\mu$rad and $g = 9.816(1)$ m/s$^2$. This indicates that the determination of $\theta_0$ using the tilt-scanned fringe is significantly less dependent on the exact value of $g$ compared to the maxima-search method.
In the above analysis, it is assumed that variations in the tilt of the Raman lasers have no influence on the interferometer except varying $\vec{k}_{\rm{eff}}\cdot\vec{g}$. This assumption is reasonable, as $k_{\rm{eff}} g T^2$ is sufficiently large, making the effect of tilt variations on $\vec{k}_{\rm{eff}} \cdot \vec{g}$ dominant. The sensitivity of the tilt-scanned fringe to $\theta_0$ can also be interpreted from another perspective: it tells that different fringes are sensitive to different quantities by comparing Figs. \ref{fringe} (a) and \ref{fringe} (b). The phase-scanned fringe interrogates the phase, making it primarily sensitive to phase shifts, whereas the tilt-scanned fringe directly interrogates the tilt, making it inherently sensitive $\theta_0$.

\begin{figure}
    \centering
    \includegraphics[width=0.75\linewidth]{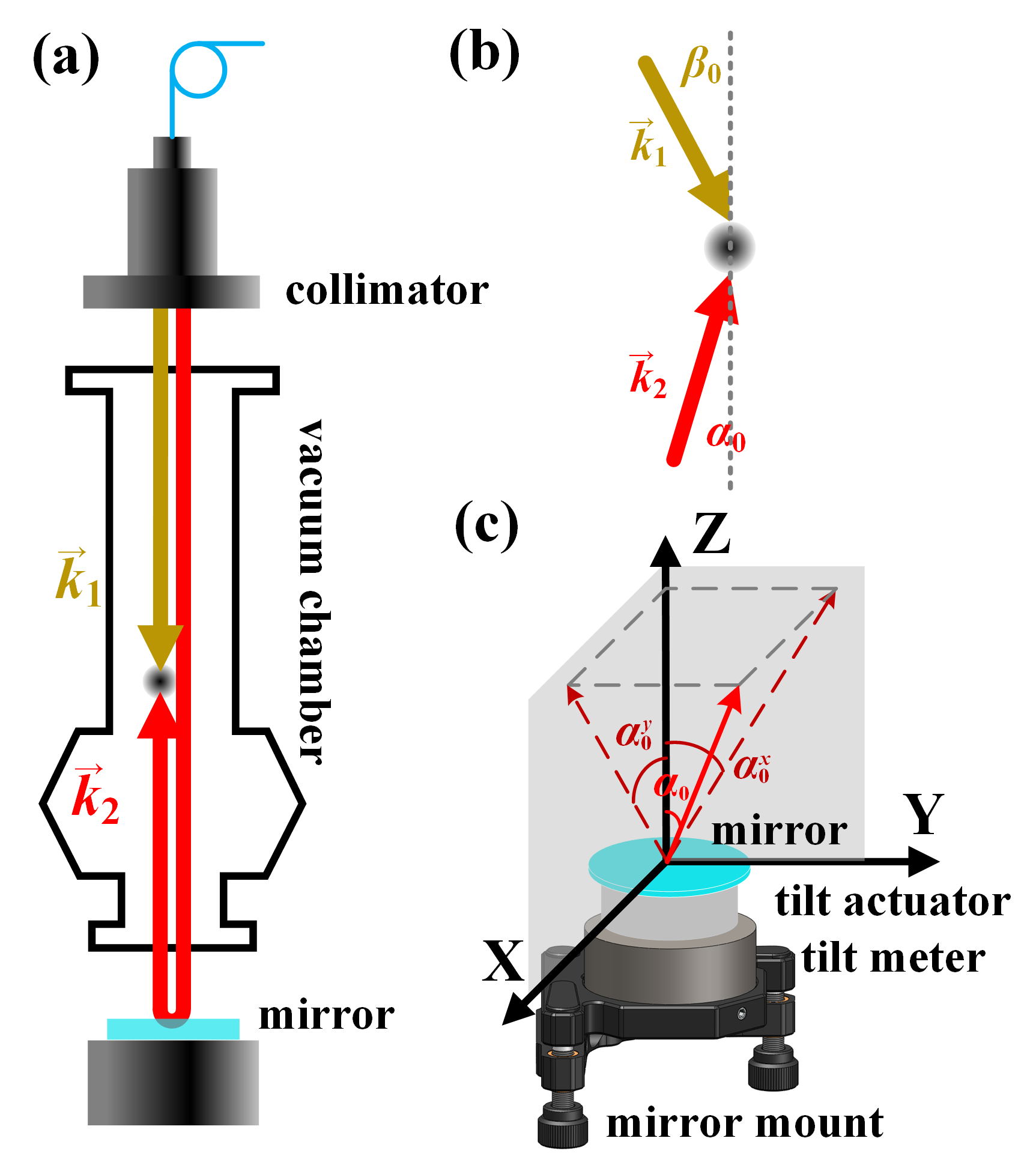}
    \caption{Experimental setup: (a) Retro-reflecting configuration for the Raman lasers. The Raman beams are collimated using a beam expander before propagating downward and being retro-reflected by a mirror positioned beneath the vacuum chamber. For the demonstration of the tilt determination, atoms interacts with the counter-propagating beams pair consisting of downward-propagating $\vec{k}_1$ beam and upward-propagating $\vec{k}_2$ beam.
    (b) The tilts of the $\vec{k}_1$ and $\vec{k}_2$ beams relative to the vertical direction are denoted as \(\beta_0\) and \(\alpha_0\). The angle \(\beta_0\) is primarily determined by the alignment of the beam expander, while \(\alpha_0\) is influenced by both the beam expander alignment and the orientation of the mirror. 
    (c) For precise alignment, the mirror is mounted on a tilt actuator capable of adjusting the beam direction along the X and Y axes via voltage control. The tilt actuator is installed on a mirror mount, which provides coarse adjustments and allows modulation of \(\alpha_0\) along two dimensions. A tilt meter is also present to  monitor changes in the stage's tilt.}
    \label{setup}
\end{figure}
\section{Experiment}
We demonstrate the proposed method for determining the tilt of the Raman lasers using the atom gravimeter HUST-QG developed by our team \cite{Xu_2022_evaluation}. The HUST-QG is a transportable instrument that participated in the tenth International Comparison of Absolute Gravimeters (ICAG) \cite{wu2021results}.  
In brief, approximately \(10^8\) \(^{87}\)Rb atoms are launched from a three-dimensional magneto-optical trap (3D-MOT) with an initial velocity of 3.45 m/s. During the launch, the atom cloud is further cooled to approximately 4 \(\mu\)K using moving molasses. After a flight time of 130 ms, the atom cloud enters the interferometer tube, where it undergoes state preparation via a Raman \(\pi\) pulse, followed by an interferometric sequence consisting of $\pi$/2-$\pi$-$\pi$/2 Raman pulses. The interrogation time \(T\) between the interferometric pulses is 200 ms.  
After completing the interferometric process, the atoms are detected using normalized fluorescence detection as they return to the detection chamber. The entire measurement cycle, as described above, takes 1 s per shot. The gravimeter HUST-QG demonstrates a typical short-term sensitivity of 24 \(\mu\)Gal/Hz\(^{1/2}\) in a quiet environment, with a Type B uncertainty of 3 \(\mu\)Gal \cite{Xu_2022_evaluation}.

\begin{figure}
    \centering
    \includegraphics[width=1\linewidth]{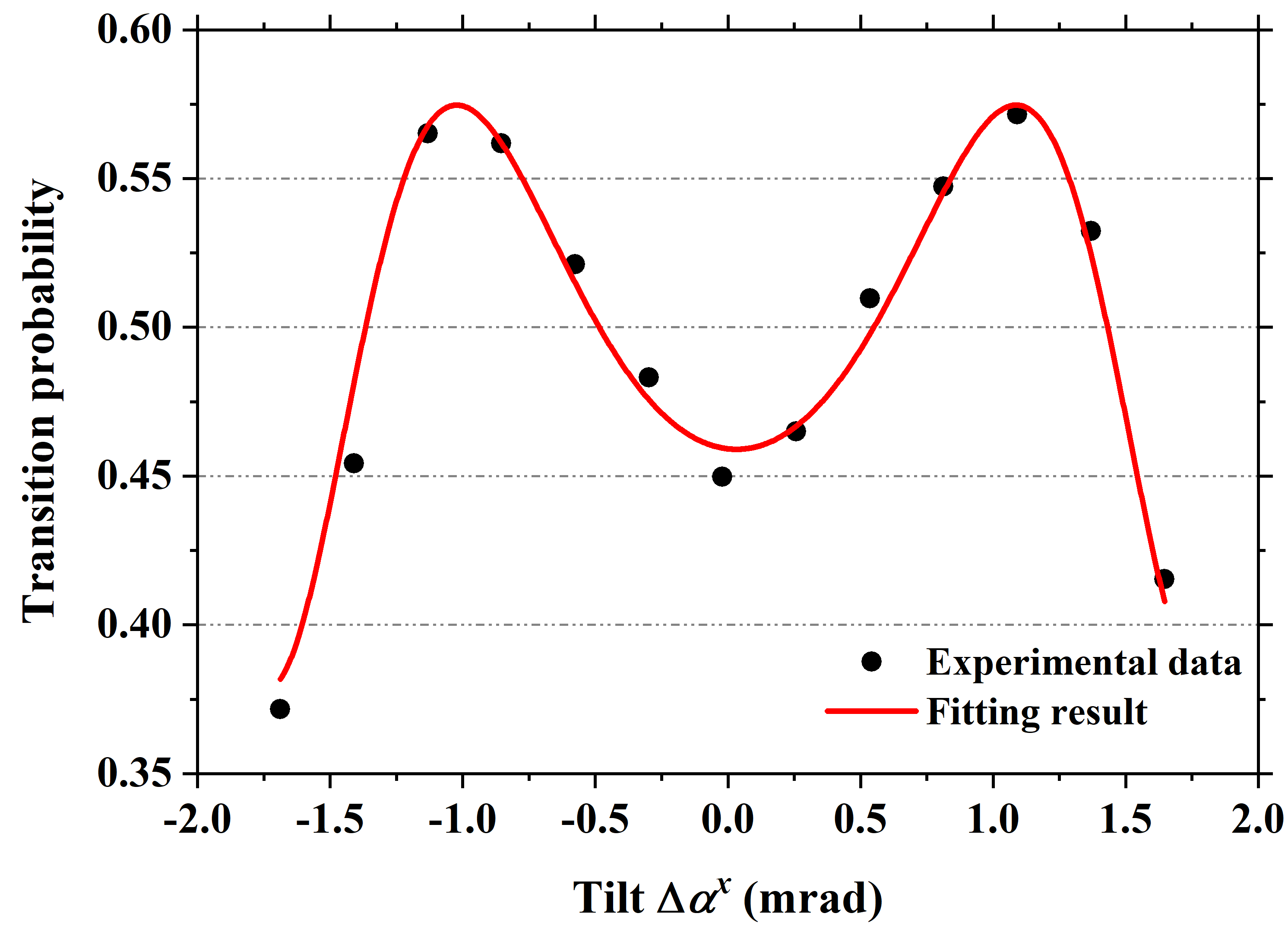}
    \caption{A typical obtained tilt-scanned fringe.Each data point corresponds to a shot measurement with a time of 1 s. Thus one tilt-scanned fringe with a measurement cycle time of 13 s can yield a determination of the tilt.}
    \label{fringedata}
\end{figure}

The Raman lasers in the gravimeter HUST-QG are configured in the commonly used retro-reflecting arrangement. As illustrated in Fig. \ref{setup}, the Raman lasers consists of two coherent beams that propagate downward after collimation. These beams originate from two diode lasers, which are phase-locked via an optical phase-locked loop (OPLL) within the HUST-QG setup. The frequency difference between the two beams is approximately 6.834 GHz, corresponding to the clock transition frequency between the two hyperfine ground states of \(^{87}\)Rb atoms.  
The downward-propagating beams are retro-reflected by a mirror positioned at the bottom of the gravimeter. Denoting the wave vector of the higher-frequency beam as \(\vec{k}_1\) and that of the lower-frequency beam as \(\vec{k}_2\), the effective wave vector of the Raman lasers \(\vec{k}_{\rm{eff}}\) is defined as \(\vec{k}_{\rm{eff}} = \vec{k}_1 - \vec{k}_2\). In the retro-reflecting configuration, there are two counter-propagating beam pairs with opposite directions of \(\vec{k}_{\rm{eff}}\), enabling two-photon stimulated Raman transitions essential for gravity measurements. For clarity in the subsequent discussion, we use the pair consisting of the downward-propagating \(\vec{k}_1\) beam and the upward-propagating \(\vec{k}_2\) beam as an example.

The collimator and retro-reflecting mirror are mounted on adjustable holders, allowing the directions of the downward-propagating and upward-propagating beams to be independently fine-tuned using adjustment screws. Initially, the directions of both laser beams, \(\vec{k}_1\) and \(\vec{k}_2\), are aligned to within 1 mrad of the vertical direction, using a liquid surface as a reference. After this initial alignment, the tilt of the Raman lasers must be measured to facilitate any further necessary adjustments. In general, the tilts of \(\vec{k}_1\) and \(\vec{k}_2\) can differ, and these are represented as \(\beta_0\) and \(\alpha_0\), respectively, as illustrated in Fig. \ref{setup}.

For this demonstration, only the determination of \(\alpha_0\) is performed. To measure the tilt, its two projections along two orthogonal horizontal rotational axes, as illustrated in Fig. \ref{setup}(c), must be determined. These projections are denoted as \(\alpha_0^x\) and \(\alpha_0^y\), respectively. Once the projections are obtained, the tilt \(\alpha_0\) can be calculated using the following expression 
\begin{equation}
    \alpha_0=\sqrt{(\alpha_0^x)^2+(\alpha_0^y)^2},
\label{theta0}\end{equation}
where the condition \(|\alpha_0| < 1\) mrad is assumed. The procedures for measuring \(\alpha_0^x\) and \(\alpha_0^y\) are identical; however, for simplicity, only the projection along one axis is measured in this demonstration. This simplified approach is expected to clearly and concisely illustrate the validity of the proposed method.
Thus, Eq. (\ref{Probability_theta}) can be rewritten as 
\begin{equation}
    P=[1-\cos{(k_2 g T^2\cos{\alpha_0^y}\cos{(\alpha_0^x+\Delta \alpha^x)}+\Delta \phi')}]/2,
\label{Probability_theta_beta}\end{equation}
where an approximation of Eq. (\ref{theta0}) is substituted. In comparison with Eq. (\ref{Probability_theta}), $\Delta \phi'= k_1 g T^2\cos{\beta_0}+\Delta \varphi$, including the invariant part of $\vec{k}_{\rm{eff}}\cdot \vec{g}T^2$ for the tilt-scanned fringe.
\begin{figure}
    \centering
    \includegraphics[width=1\linewidth]{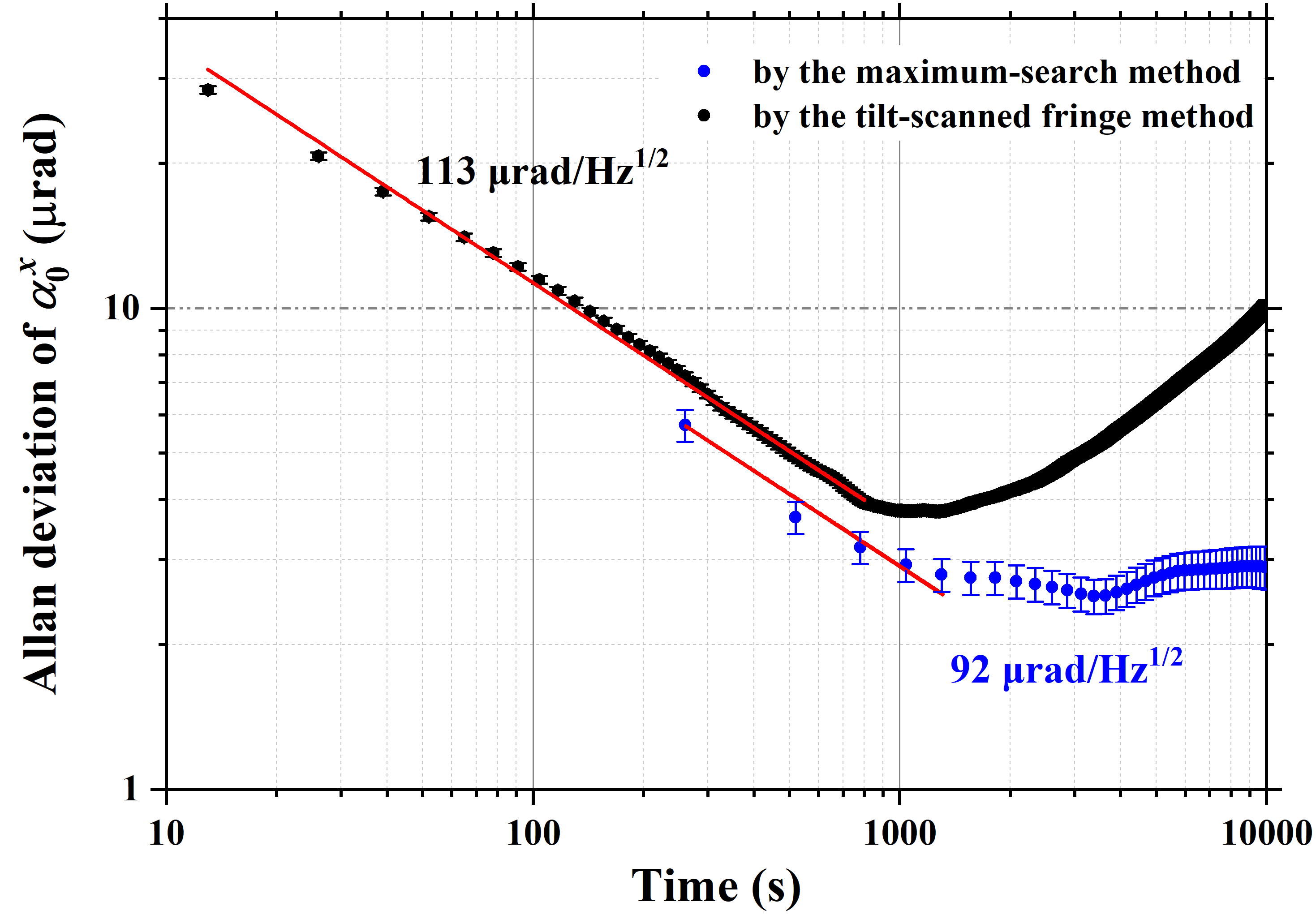}
    \caption{Allan deviations obtained from continuous measurement data for the two methods for determining the tilt of the Raman lasers. The results shows that the tilt-scanned fringe method can achieve a short-term sensitivity of 113 $\mu$rad/Hz$^{1/2}$, comparable to the sensitivity of 92 $\mu$rad/Hz$^{1/2}$ for the usual maximum-search method. However, the measurement cycle time is only 13 s for the tilt-scanned fringe method, manifesting an order of magnitude improvement over the usual maximum-search method.}
    \label{AllanDeviation}
\end{figure}

\begin{figure}
    \centering
    \includegraphics[width=1\linewidth]{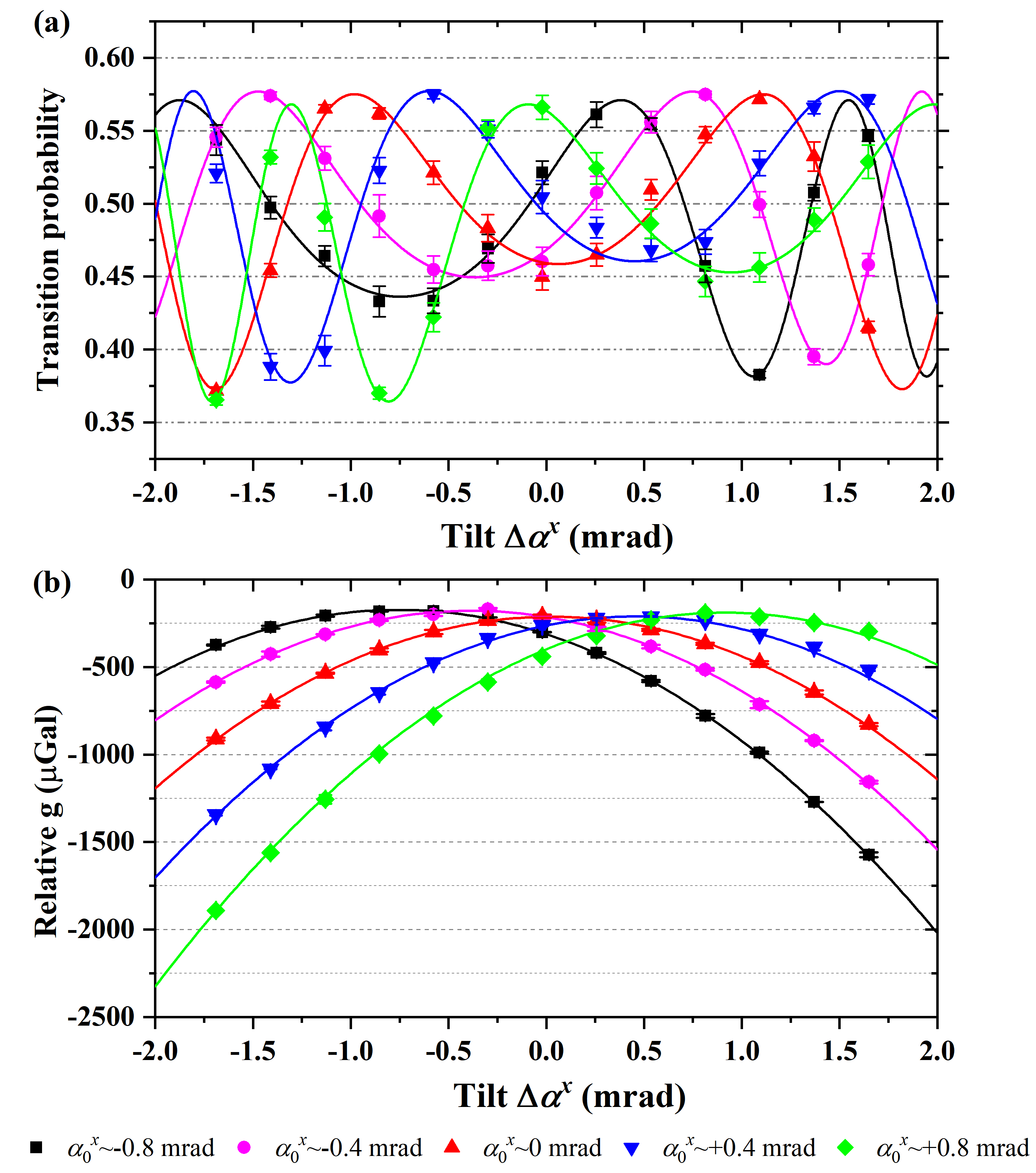}
    \caption{
    Measurements of different $\alpha_0^x$ using the two methods. $\alpha_0^x$ is changed in five steps using the coarse adjustment of the mirror mount shown in Fig. \ref{setup}(c) by an increment of about 400 $\mu$rad (indicated by the tilt meter). (a) Tilt-scanned fringes for the measurements with each fringe corresponding to one value of $\alpha_0^x$. For each fringe, $\Delta\alpha^x$ is varied from -1.68 mrad to +1.68 mrad by the tilt actuator to scan the fringe in 13 steps, where ten shots measurements are repeated for the average for each step. (b) Parabolic variations of the $\vec{g}$ projection along with the tilt. Each parabolic variation corresponds to one value of $\alpha_0^x$ with $\Delta\alpha^x$ also varied from -1.68 mrad to +1.68 mrad. Each data displayed for the parabolic variations represents an average of measurement results of three phase-scanned fringes with each fringe taking twenty shots measurements. The lines in (a) and (b) represents respective fitted lines.
    }\label{theta0variation}
\end{figure}

A voltage-controlled tilt actuator (PI S-340.ASL) is employed to adjust the orientation of the retro-reflecting mirror, while a tilt meter mounted on the rotational motion stage is used to record the mirror's posture. Both the tilt actuator and tilt meter operate in two dimensions, with their axes aligned correspondingly. The X and Y axes shown in Fig. \ref{setup} (c) are defined by the rotational axes of the tilt actuator. 
For the measurement of \(\alpha_0^x\), the tilt of \(\vec{k}_2\) is scanned by rotating the retro-reflecting mirror along the X axis using the tilt actuator. A typical tilt-scanned fringe obtained during the experiment is shown in Fig. \ref{fringedata}, where \(\Delta\alpha^x\) varies from -1.68 mrad to 1.68 mrad in steps of 0.28 mrad. The data are fitted using the function described earlier, yielding a result of $\alpha_0^x$=-32(16) \(\mu\)rad.  
This result demonstrates that one tilt-scanned fringe with a measurement cycle time of 13 s is indeed capable to give a determination of the tilt with high precision, as expected by the simulation above.
To assess the short-term sensitivity of the tilt projection determination using the tilt-scanned fringe method, \(\alpha_0^x\) is measured continuously while \(\alpha_0^x\) and $\alpha_0^y$ remain unchanged. Allan deviation analysis of the resulting data, as shown in Fig. \ref{AllanDeviation}, indicates a short-term sensitivity of \(113\) \(\mu\)rad/Hz\(^{1/2}\) for the proposed method.  
For comparison, the maximum-search method based on phase-scanned fringe was also used to measure the tilt projection. In this measurement, \(\Delta\alpha^x\) is also varied from -1.68 mrad to 1.68 mrad in steps of 0.28 mrad, with the effective phase of the Raman lasers scan in 20 steps to obtain a full fringe at each \(\Delta\alpha^x\). This process requires 260 seconds for a single measurement of the tilt projection. Continuous measurements using the maximum-search method are also performed, and the corresponding Allan deviation analysis shown as blue dots in Fig. \ref{AllanDeviation} reveals a short-term sensitivity of 92 \(\mu\)rad/Hz\(^{1/2}\).  
The comparison illustrates that the tilt-scanned fringe method reduces the measurement cycle time by over an order of magnitude while maintaining a short-term sensitivity comparable to that of the maximum-search method.

\begin{figure}
    \centering
    \includegraphics[width=01\linewidth]{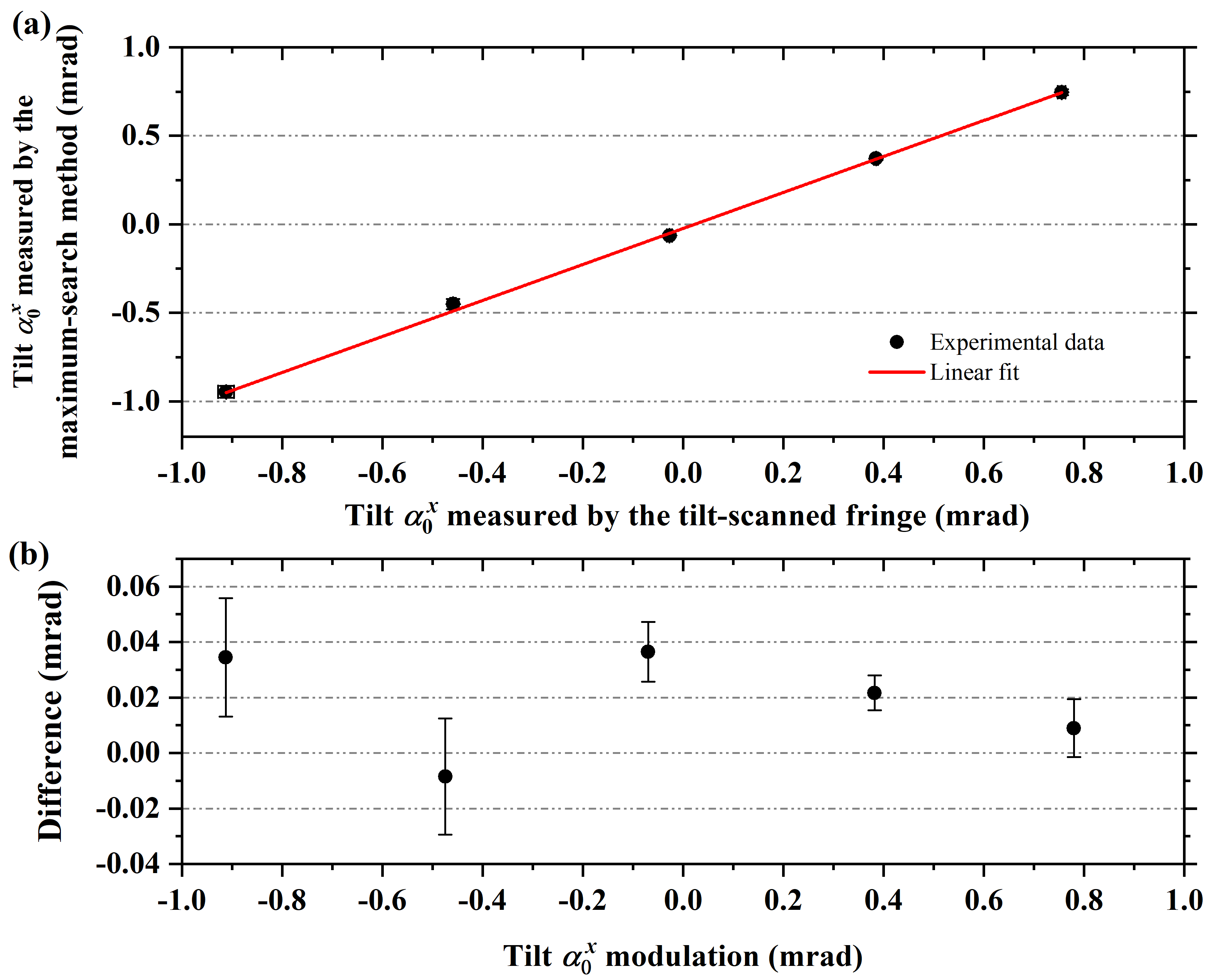}
    \caption{
    Comparison of the measurement results by the two methods. (a)The measurement result by the tilt-scanned fringe versus that by the maximum-search method. A linear fit to the data yield a slope of 1.02(2) mrad/mrad and an intercept of -23(8) $\mu$rad. (b) Differences between the measurement results by the two methods. The maximum discrepancy is 36 $\mu$rad, corresponding to an impact on gravity measurements of less than 1 $\mu$Gal.
    }
    \label{comparison}
\end{figure}

To further validate the measurement accuracy of the tilt-scanned fringe method, $\alpha_0^x$ is varied in increments of approximately 400 \(\mu\)rad using the adjustment screw of the retro-reflecting mirror mount. For each increment, $\alpha_0^x$ is measured using both the tilt-scanned fringe method and the maxima-search method. Typical tilt-scanned fringes for different values of $\alpha_0^x$ are shown in Fig. \ref{theta0variation}(a). Each displayed fringe corresponds to 13 steps of $\Delta\alpha^x$, with each step representing the average of 10 shot measurements.  
Additionally, variations in the projection of \(\vec{g}\) for different $\alpha_0^x$, measured using the phase-scanned fringe method, are shown in Fig. \ref{theta0variation}(b). Each data point in Fig. \ref{theta0variation}(b) represents the average of 3 fringes, where each fringe consists of 20 phase-scanned shot measurements.  
The measurement results obtained from the two methods are compared in Fig. \ref{comparison}, demonstrating good agreement. The maximum discrepancy between the results of the two methods is only 36 \(\mu\)rad over a measurement range from -0.91 mrad to 0.78 mrad. This comparison confirms that the proposed tilt-scanned fringe method is sufficiently accurate for current \(\mu\)Gal-level gravity measurements using atom gravimeters.  

\section{Conclusion}
The tilt of \(\vec{k}_1\) can be determined once both \(\alpha_0^x\) and \(\alpha_0^y\) are measured. For instance, the direction of \(\vec{k}_2\) can be adjusted along the X and Y axes using the tilt actuator until the laser power re-injected into the fiber coupler of the Raman laser collimator from the retro-reflected beam reaches its maximum. At this point, the direction of \(\vec{k}_2\) can be assumed to coincide with that of \(\vec{k}_1\), with negligible deviation.  
The tilt variation of \(\vec{k}_2\) during this adjustment process can be inferred from changes in the voltage applied to the tilt actuator. 
Denoting these variations as \(\delta\alpha^x\) and \(\delta\alpha^y\) along the corresponding axes, the tilt of \(\vec{k}_1\) can be determined as \(\beta_0^{x,y} = \alpha_0^{x,y} - \delta\alpha^{x,y}\). In certain atom gravimeters \cite{zhang2021carbased,Bodart2010pyramidal}, the tilt of the entire instrument can be adjusted such that the directions of \(\vec{k}_1\) and \(\vec{k}_2\) remain aligned at all times. In such cases, determining the tilt of \(\vec{k}_{\rm{eff}}\) becomes significantly simpler using the tilt-scanned fringe method.

In conclusion, we propose a tilt-scanned fringe method for fast and precise of the determination of the tilt of Raman lasers in atom gravimeters. This method involves varying the tilt of the Raman lasers to generate fringes. The resulting tilt-scanned fringe is highly sensitive to the laser tilt while being significantly less influenced by the precise value of gravitational acceleration compared to the conventional maxima-search method. 
We have successfully demonstrated the proposed method using our developed atom gravimeter. The results indicate that the tilt-scanned fringe method reduces the measurement cycle time by approximately an order of magnitude while maintaining comparable short-term sensitivity and accuracy to the maxima-search method. This approach is expected to simplify and accelerate the deployment of atom gravimeters for gravity surveys.

\begin{acknowledgments}

\smallskip
We thanks Prof. Minkang Zhou and Lushuai Cao for the meaningful discussions regarding this article. This work was supported by the National Natural Science Foundation of China (Grants No. 12205110, No. U2341247, No. 12474482). 

\end{acknowledgments}

\bibliography{sample}

\end{document}